\documentclass[aps,prc,showpacs,twocolumn,preprintnumbers,floatfix,
showkeys,tightenlines]{revtex4}
\usepackage{amsmath,amssymb,amsfonts}
\usepackage{graphicx}
\usepackage{color}

\allowdisplaybreaks
\begin{document}
\title{
Gamma flashes from relativistic electron-positron plasma droplets 
%How bright is a electron-positron droplet?\\
}
\author{Munshi G. \surname{Mustafa}$^{1,2,}$} 
\email{musnhigolam.mustafa@saha.ac.in}
\author{B. \surname{K\"ampfer}$^{1,3,}$}
\email{b.kaempfer@fzd.de}
\affiliation{
$^1$Forschungzentrum Dresden-Rossendorf, 01314 Dresden, Germany \\ 
$^2$Theory Division, Saha Institute of Nuclear Physics, 1/AF Bidhannagar, Kolkata 700064, India\\
$^3$Institut f\"ur Theoretische Physik, TU~Dresden, 01062 Dresden, Germany}

\begin{abstract}
Ultra-intense lasers are expected to produce, in near future, 
relativistic electron-positron plasma droplets.
Considering the local photon production rate in 
complete leading order in quantum electrodynamics (QED),
we point out that these droplets are interesting
sources of gamma ray flashes.
  
\end{abstract}
%\preprint{SINP/TNP/06-30}
\maketitle

Fastly progressing high-intensity laser technology \cite{Tajima_Mourou,Accelerator}
offers the perspective to create relativistic plasmas composed of
electrons ($e^-$), positrons ($e^+$) and photons ($\gamma$) under
laboratory conditions. Accordingly simulations have been performed,
{\it e.g.}, in \cite{Shen_MtV,Liang} for special scenarios. Ideally would be the
creation of an equilibrated $e^+e^-\gamma$ plasma of
sufficient size exceeding the mean free path of photons. Such a
state of matter with properties reviewed in \cite{Thoma} is governed by
thermo-field theory applied to QED and also in \cite{Remo} by kinetic theory. 
If such a plasma could be
produced by table-top laser installations, it would represent the
QED pendent to the QCD quark-gluon plasma currently investigated
with large-scale accelerators \cite{QGP_review}.
%and also to the early Universe~\cite{Early_U},
% active Galactic nuclei~\cite{Galactic} and gamma-ray brusts~\cite{GBR}. 
Relativistic plasmas
have some interesting specific features, such as collective plasmon
or plasmino excitations \cite{plasmon_plasmino} with accompanying van
Hove singularities \cite{Peshier_Thoma}. Electron-positron plasmas are also
interesting for astrophysical scenario~\cite{Takabe}. Temperatures of about 
$T \sim 10$ MeV, in a laser-generated plasma,
would open, furthermore, channels for muon or pion production
\cite{Rafelski}.  While these channels are interesting for their own right (see
\cite{Keitel} for another avenue for laser driven particle production),
a hot $e^+e^-\gamma$ plasma droplet, exploding after the creation
process by the enormous thermodynamic pressure, is a source of
$\gamma$ radiation. Such short $\gamma$ flashes may be of future use
in ultra-fast spectroscopic investigations of various kind.

In this note, we are going to present estimates of the 
photon spectrum emerging from a $e^+e^-\gamma$ plasma droplet with
temperatures in the $10$ MeV range. In contrast to often employed
particle-in-cell (PIC) 
%\marginpar{PIC}
simulations (see \cite{Shen_MtV} for a study of the start-up phase of a 
similarly hot plasma), we base our
considerations on results from QED thermo-field theory. 
The expansion dynamics is treated in a schematic way, as our
emphasis is on the  photon emission characteristics. 
%As initial state
%we assume an equilibrated $e^+e^-\gamma$ droplet of $T = 10$ MeV 
%and a radius of 2 $n$m, as assumed in \cite{Rafelski}. 
%as in \cite{Thoma}, and also mention results 
%Later, we briefly discuss the
%consequences of an asymmetry of $e^-$ and $e^+$ densities by
%including a non-zero chemical potential.
%and effects of chemical undersaturation by means of fugacities.

The calculation of the local spontaneous photon emission rate  
%with energy $E_\gamma > T$
from a QED (electron-positron) plasma has been outlined in \cite{Moore}
to  complete leading order in 
electromagnetic coupling $\alpha$ by including two-loop order.
The result, at the heart of our note, may be presented as
\begin{eqnarray}
\frac{dN}{d^4x d^3E_\gamma} &=& 
\frac{2\alpha m^2_\infty n_F(E_\gamma)}{(2\pi)^3\ E_\gamma} 
\left [\ln \left (\frac{T}{m_\infty}\right )+\frac{1}{2} \ln \left ( \frac{2E_\gamma}{T} \right )
\right .  \nonumber  \\
&+&\left . C_{22}\left(\frac{E_\gamma}{T}\right) \ 
+  \ C_{\rm{b}}\left(\frac{E_\gamma}{T}\right) 
+  C_{\rm{a}}\left ( \frac{E_\gamma}{T}\right) \right ] \label{local}
\end{eqnarray}
where $n_F(E_\gamma)=(\exp (E_\gamma/T)+1)^{-1}$ denotes the Fermi distribution, 
and the dynamically generated asymptotic mass squared of the electron is 
$ m^2_\infty = 2 m^2_{\rm{th}}$.
The thermal mass squared is given by $m^2_{\rm{th}}=e^2T^2/8$ 
with $e^2=4\pi\alpha$  
(The considerations apply for $T > m_{e^\pm}$ which are different from the non-relativistic
case where the relevant scales are given by the masses of plasma particles and the temperature~\cite{Thoma}.)
$E_\gamma = u \cdot k$ is a short hand notation for the Lorentz scalar product
of the medium's four-velocity $u(t, \vec x)$ and the photon's four-momentum $k$. 
This rate includes $2\leftrightarrow2$ processes from one loop~\cite{Kapusta}, 
{\it viz.}, Compton scattering and pair annihilation which generate the leading logarithmic 
contribution (first three terms in (\ref{local})).
In addition, the rate also includes the inelastic processes~\cite{Aurenche} from two-loop like 
bremsstrahlung (fourth term) and off-shell pair annihilation (fifth term) with the 
correct incorporation~\cite{Moore,Moore1} of the Landau-Pomeranchuk-Migdal
effect that limits the
coherence length of the emitted radiation.  A common feature of the latter two processes is that 
they have off-shell fermion next to the vertex where the photon is emitted, and the virtuality 
of the fermion becomes very small if the photon is emitted in forward direction. 
However, contrary
to the one-loop diagrams, the singularity is linear instead of logarithmic, 
and it brings a factor
$T^2/m^2_{\rm {th}}$ making it the same order in $\alpha$ as that of one-loop. 
The functions $C$s are independent of $\alpha$
(because the relative importance of scattering with a photon to an another 
charged particle in the
plasma is essentially given by the ratio $m^2_\infty/m^2_D$, where $m_D$ is the Debye screening mass) 
but are nontrivial functions of $y=E_\gamma/T$ determined by a set of integral 
equations~\cite{Moore,Moore1}.  The numerical results for electron-positron plasma
are described quite accurately by phenomenological fits
in the domain $0.1 \leq y \leq 30$ which is sufficient for our consideration.
The $C_{22}$ is same in
QED and QCD because
the interference effect cancels out %~\cite{private} 
and is given as
%~\cite{Moore}
\begin{equation}
C_{22}(y) \simeq 0.041y^{-1}-0.3615+1.01e^{-1.35y} , \label{c22}
\end{equation}
whereas the other two coefficients due to inelastic processes 
in QED are fitted together yielding
%to a single expression as~\cite{private}
\begin{equation}
C_{\rm{b}}(y)+ C_{\rm{a}}(y)\simeq \frac{2 a \ln(b+1/y)}{y^c}+\frac{2d y}{\sqrt{1+y/f}}, 
\label{cba}
\end{equation}
where $a=0.374958$, $b=0.431304$, $c=-0.05465$, $d=0.157472$ and $f=1.73085$. 

\begin{figure}[!htbp]
\vspace{-0.4cm}
\begin{center}
%\showthe\columnwidth % Use this to determine the width of the figure.
%\includegraphics[width=\columnwidth,keepaspectratio]{static_qed.ps}
\includegraphics[width=\columnwidth,height=6.5cm]{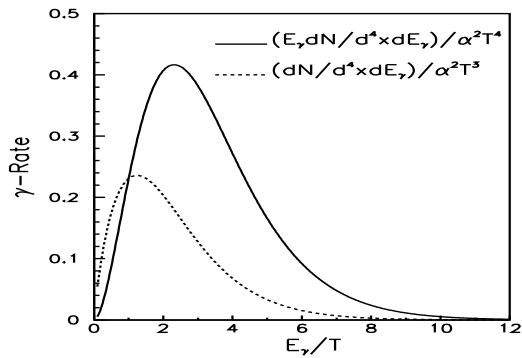}
\vspace{-2.0cm}
\caption{\label{fig1} Differential photon emission rate (dashed) as well that of energy 
weighted (solid)  for an equilibrated, symmetric $e^+ e^- \gamma$ plasma as a function of 
photon energy scaled with temperature. 
} 
\end{center}  
\end{figure}

In Fig.~\ref{fig1}, the
scaled differential photon rate 
%with temperature $T$ in (\ref{local}) 
is displayed as a
function of scaled photon energy.
%, $E_\gamma/T$. 
A comparison reveals that 
the low-energy photon rate is
suppressed in the QED plasma relative to a QCD plasma (consisting of quarks, anti-quarks
and gluons, all strongly interacting). 
The reason is that a photon has a non-negligible medium
induced thermal mass, while in QCD its thermal mass is suppressed 
relative to the quarks by $(e^2/g^2)$,
where $g$ is the strong coupling in QCD.

The damping of a photon is related to the absorption of it in the medium. The damping rate  
can be obtained from the photon production rate in (\ref{local}) by the principle of detailed
balance~\cite{Moore,Weldon,Thoma1} as
\begin{equation}
\Gamma_\gamma(E_\gamma) = \frac{(2\pi)^3}{4} \ e^{E_\gamma/T} \ \frac{dN}{d^4xd^3E_\gamma} \ ,
\label{damp}
\end{equation}
where the prefactor, $(2\pi)^3/4$, is a matter of definition~\cite{Thoma1}.
The mean free path, $\lambda_\gamma$, 
of a photon in a medium is inversely related to the photon 
damping rate as 
%\begin{equation}
$\lambda_\gamma ={1/\Gamma_\gamma}$. % \ . \label{fpath} 
%\end{equation}
\begin{figure}[!htbp]
\vspace{-0.4cm}
\begin{center}
%\showthe\columnwidth % Use this to determine the width of the figure.
%\includegraphics[width=\columnwidth,keepaspectratio]{fpath_static_qed_e.ps}
\includegraphics[width=\columnwidth,height=6.5cm]{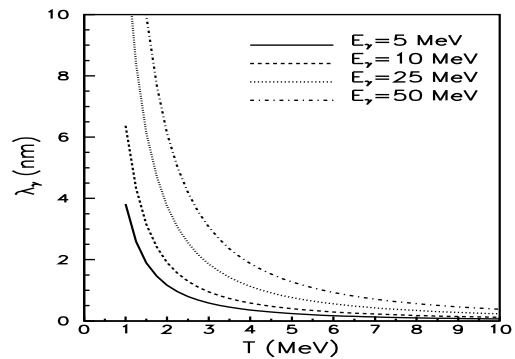}
\vspace{-2.0cm}
\caption{\label{fig2}  
Mean free path of photons of energy $E_\gamma$ 
in an equilibrated, symmetric $e^+ e^- \gamma$ plasma 
as a function of temperature.
%for a wide range of photon energies $E_\gamma$. 
} 
\end{center}  
\end{figure}
In Fig.~\ref{fig2} the mean free paths of photons are displayed as a function
of temperature of the electron-positron plasma for a wide range of photon energies.
This figure reveals very interesting features of the $e^+e^- \gamma$  plasma droplet. 
Apparently at initial temperature, it appears to be completely opaque as discussed in
Refs.~\cite{Thoma,Rafelski}. 
However, it may not be so forever as we will see below.
For a given photon energy $E_\gamma$ the mean free path increases with decreasing 
temperature implying a dilute system after some time. 
On the other hand, for a given $T$ it  increases 
with energy, $E_\gamma$, making the system  more transparent to the low energy
photons depending upon the size of the plasma droplet. 
However, the realistic scenario would be to consider the expansion of 
the droplet due to thermodynamic pressure. 

We consider the expansion 
with spherical geometry of radius $R(t)=R_0+v(t)t$, where
$R_0$ is the initial size of the droplet, $v(t)$ 
is the surface velocity of the droplet 
assumed as $v(t)=\eta t$, where $\eta$ is a constant 
having dimension $c^2/[{\rm {Length}}]$.
We adopt a linear flow profile $v(t,r)= v(t) r /R(t)$. 
The space-time integrated radiation spectrum follows from
\begin{eqnarray}
E_\gamma \frac{dN}{dE_\gamma} &=& 2\int_0^{t} dt 
\int_0^{R(t)} dr \, r^2 \int_0^{2 \pi} d\phi \ \nonumber \\
&&\times \ 
\left ( E_\gamma \frac{dN}{d^4x \, dE_\gamma} \right ) 
\Theta \left [\lambda_\gamma(t) - L(r,\phi)\right ],
\label{spectra}
\end{eqnarray}
where $L(r,\phi) = \left (R^2(t)-r^2 \sin^2\phi\right )^{1/2}-r\cos\phi$ 
is the distance~\cite{Muller} the 
photon travels after it is created at a point $r$ with an angle $\phi$
relative to the radial direction.
This expression takes into account that the radiation can only be emitted
from such positions $r$ and in such directions $\phi$ where the
distances to the surface $R(t)$ is shorter than the mean free path $\lambda_\gamma(t)$.
The temperature at a time instant $t$ is calculated 
by employing the energy conservation relation:
$E(t)=E_0-E^{\rm{em}}_\gamma(t)-E^{\rm{kin}}(t)$, 
where $E_0$ is the initial energy of the droplet
obtained from the total initial energy density, 
$ \epsilon(T_0)=11\pi^2T^4_0/60\ \simeq 3.8 \times 10^{29}$ Jm$^{-3}$ at $T_0=10$ MeV.
We note that $E_0 \ (\propto T^4_0R_0^3 \ )$  depends crucially on the initial droplet size.
$E_0 \simeq 13$ kJ for a droplet of radius $R_0=2 \ n$m with the same reasoning as in 
Ref.~\cite{Rafelski}.
%($13\times 10^{9}$ kJ for $R_0=2\ \mu$m ). 
(It is an interesting and challenging topic how 
the laser pulse energy can be deposited in a smallest possible spatial and temporal volume.
The scenario in Ref.~\cite{Shen_MtV,Rafelski} envisages, {\textit{e.g.},} 
an initial stage of compressed foil with $(1-2) \ n$m thickness and $T_0\sim 10$ MeV).
Now, at any instant of time $t$,  
one can obtain the energy of the emitted photon as
$E^{\rm {em}}_\gamma(t)=\int dE_\gamma \ \left (E_\gamma dN/dE_\gamma \right )$ 
from (\ref{spectra}), 
and the kinetic energy of the droplet as
$E^{\rm{kin}} (t) = 4\pi \int \ dr \ r^2 \ \epsilon(T(t)) \gamma v(t,r)$, 
with $\gamma=1/(1-v^2(t,r)/c^2)^{1/2}$.
The evolution is truncated at  time $t_f$ at which $T$ cools down to $1$ MeV.
For lower temperatures, the electron rest mass becomes a relevant scale which
modifies the employed rate. As we are interested in the hard part of the emission
spectrum, $E_\gamma \ge 1$ MeV, the late evolution with $T \le 1$ MeV is not 
essential for our purposes.

\begin{figure}[!htbp]
\vspace{-0.4cm}
\begin{center}
%\showthe\columnwidth % Use this to determine the width of the figure.
\includegraphics[width=\columnwidth,height=6.5cm]{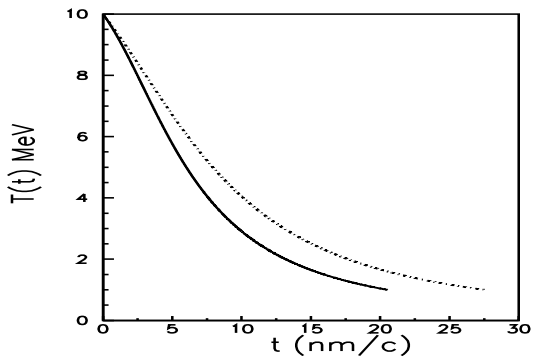}
\vspace{-2.0cm}
\caption{\label{fig3} Temperature as a function of time for $\eta = 0.03 \ c^2/n$m (solid) 
and $0.01\ c^2/n$m (dashed).  %with the initial droplet size $R_0=2\ n$m.
} 
\end{center}  
\end{figure} 

In Fig.~\ref{fig3}, the evolution of $T(t)$ is displayed for two surface velocities. 
%for a droplet with initial radius $2 \ n$m.  
The expansion lasts up to $t_f\sim (20 \ -30)\ n$m/$c$.
A realistic three-dimensional expansion dynamics can be obtained from relativistic
hydrodynamics from given initial conditions, in particular phase space distribution.
This deserves separate dedicated investigations which we postpone.
A relevant scale is the velocity of sound ($1/\sqrt{3}$ in an ideal relativistic plasma)
at which a rarefaction wave travels from the surface towards the center is compatible with
our consideration.
%The averaged overall expansion velocity, especially for later times, may be
%significantly smaller.      
We note that the evolution depends strongly on the system size but 
weakly on the surface expansion velocity.

\begin{figure}[!htbp]
\vspace{-0.5cm}
\begin{center}
%\showthe\columnwidth % Use this to determine the width of the figure.
\includegraphics[width=\columnwidth,height=6.5cm]{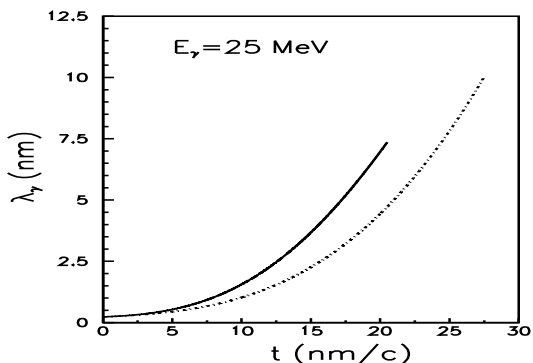}
\vspace{-2.0cm}
\caption{\label{fig4} Mean free path of a photon with $E_\gamma = 25$ MeV
as a function of time 
for $\eta = 0.03 \ c^2/n$m (solid) 
and $0.01\ c^2/n$m (dashed). 
} 
\end{center}  
\end{figure} 

In Fig.~\ref{fig4}, we display the evolution of $\lambda_\gamma(t)$ with same
initial conditions as above.
The mean free path of the photon for the expanding droplet depends on the velocity profile. 
This in turn determines the region  from which the photons may escape. The
system becomes more dilute and transparent in later time.
%It is obvious that it would be so from a thin surface layer of only a few $n$m 
%of the $e^+ e^- \gamma$ droplet 
%irrespective of its initial size. 
Inspection of Fig.~\ref{fig2} also unravels that
harder photons have a larger optical depth, as mentioned above.

\begin{figure}[!htbp]
%\vspace{-0.4cm}
\begin{center}
%\showthe\columnwidth % Use this to determine the width of the figure.
\includegraphics[width=\columnwidth,height=6.5cm]{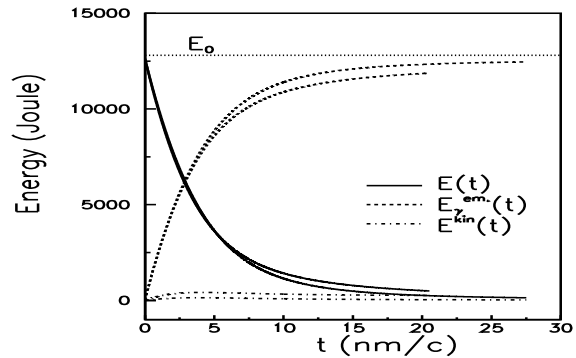}
\vspace{-2.0cm}
\caption{\label{fig5} Evolution of  $E(t)$ (solid), $E^{\rm{em}}_\gamma(t)$ (dashed) and 
$E^{\rm{kin}}(t)$ (dash-dotted)  
for $\eta = 0.03 \ c^2/n$m  and $0.01\ c^2/n$m.
A larger value of $t_f$ corresponds to a smaller value of $\eta$. 
} 
\end{center}  
\end{figure} 

In Fig.~\ref{fig5}, we exhibit the evolution of various energies with same initial conditions 
 as above. 
Initially, $E^{\rm{kin}}$ increases
with the increase of time but then decreases as the size of the system becomes larger and
the local energy density drops.
The integrated energy of emitted photons increases with time indicating 
that the most of the system's energy is attributed
to photon production. Thus, the total energy remaining in the droplet decreases with time.
The surprising fact is that cooling by photon emission is very efficient,
i.e.\  the initial energy of the plasma droplet is converted essentially into hard 
photon radiation. This limits the time span the droplet spends in a hot stage.
In addition, $e^+$ and $e^-$ leakage reduces the life time of the hot era further.
The droplet, after equilibration, appears a gamma flash. Of course,
there are relaxation processes, like heat conductivity etc.
The present estimate assumes that these are short compared with the
expansion time scale. 

\begin{figure}[!htbp]
\vspace{-0.4cm}
\begin{center}
%\showthe\columnwidth % Use this to determine the width of the figure.
\includegraphics[width=\columnwidth,height=6.5cm]{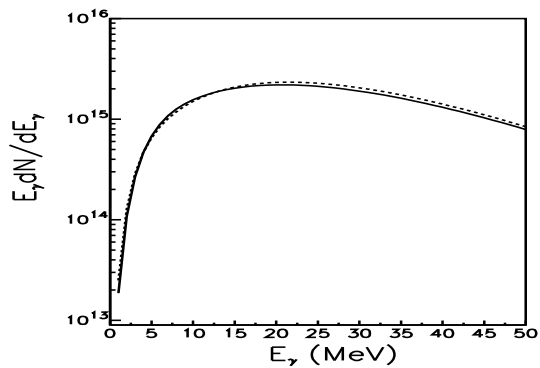}
\vspace{-2.0cm}
\caption{\label{fig6} Energy weighted photon spectrum as a function of photon
energy for surface expansion velocities with  $\eta = 0.03 \ c^2/n$m (solid) 
and $0.01\ c^2/n$m (dashed). Radiation from the ignition stage is not included.
%for the initial droplet size $R_0=2\ n$m.
} 
\end{center}  
\end{figure} 

In Fig.~\ref{fig6}, we display the energy weighted photon spectra. 
%for two expansion velocity profiles with the initial radius as before. 
As seen there is an enormous amount of photon production
from the electron-positron plasma droplet.
Note that the time integrated photon spectrum scales roughly with the initial
volume or energy $E_0$, respectively. For instance, a
one-order change in initial radius would correspond
to a three-orders of magnitude change in photon production.
%If a initial droplet of radius $2 \ n$m 
%({\textit {i.e.}}, 3 order smaller than the one considered here) is 
%expected to be formed with the same initial $T$ as discussed in Ref.~\cite{Rafelski}, 
%the $t_f$ reduces by the same order whereas the various energies in Fig.~\ref{fig5} and the 
%photon production rate in Fig.~\ref{fig6} are smaller by nine order of magnitudes due to
%the reduction in volume.

We now briefly discuss an asymmetric plasma with non-zero net electric charge. 
The chemical potential $\mu$ steers such an asymmetry.
The ratio of electron to positron densities is approximately $\propto \exp\{ 2 \mu /T\}$,
while the net density behaves like $\propto \mu T^2$ in leading order. 
That means, a noticeable difference in $e^+$ and $e^-$ densities translates into
$\mu \ge T$. 
$\mu$ enters the photon rate 
in (\ref{local}) through $m^2_\infty= e^2T^2(1+\mu^2/(\pi^2T^2))/4$, 
which takes into account the 
number of charge particles available for various processes. 
Also various $C$s  
depend on $\mu$ through the ratio 
$m^2_\infty/m^2_D \sim (1+\mu^2/(\pi^2T^2))/(1+3\mu^2/(\pi^2T^2))$ 
which accounts for, as discussed earlier, 
the relative importance of scattering with a photon to a 
charge in the plasma. Therefore, it could only be important at $\mu \geq T$.

The present considerations are armed with a state-of-the art photon emissivity. The
dynamics, in contrast, is rather schematic: constant temperature over the fireball,
linear velocity profile, prescribed surface velocity and velocity profile etc.
Envisaged calculations within relativistic hydrodynamics
including gradients and details of radiation transport
will make the scheme more realistic. Clearly, chemical undersaturation~\cite{Shen_MtV,Ruffini} 
%(PIC simulation for
%the scenario in Ref.~\cite{Shen_MtV} predicts a positron density of about $0.5\%$ of the
%electron density 
(which would correspond to a decrease of the initial energy density for the
same initial temperature) 
parameterized by fugacities need to be included along the 
lines of \cite{Oleg_BK} to arrive at a detailed picture. 
%However, we note that the off-shell annihilation
%dominates at high energy than the bremstrahlung and $2\leftrightarrow 2$ processes~\cite{Moore,Aurenche}.
%For chemically undersaturation plasmas the main effect of the fugacities $(\xi_i=n_i/n^{\rm{eq}}_i$, $n_i$
%is the number density for various species $i= e \{ e^-, e^+\} \ {\rm{and}} \ \gamma$) comes 
%from the fact that the number of emitters is reduced. One can take this into account 
%through  the number of participants in the initial states such that
%the rates will depend on $\xi_i$. The Compton is $\propto \xi_e\xi_{\gamma}$, the annihilation is 
%$\propto \xi_e^2$, the bremsstrahlung is $\propto \xi_e^2$ whereas the off-shell
%annihilation is $\xi_e^3$. For undersaturated plasma the off-shell annihilation will be largely suppressed
%and the photon spectrum will be dominated by $2\leftrightarrow 2$ and the bremstrahlung processes.
%Because of this compensating effect the photon spectra is expected to be of the order of i
%equilibrium specta~\cite{Mustafa}.
The effect of impurities,
like ions or the original material where the plasma is formed from, should  be 
considered too, as other expansion patterns, {\textit{e.g},} for the geometry envisaged
in Ref.~\cite{Shen_MtV}.

In summary we point out that a hot $e^+e^-\gamma$ plasma droplet created by
ultra-intense and short laser pulses is a technologically
interesting source of $\gamma$ flashes with continuous spectrum (and not only 
a flash of $511$ keV radiation as argued in Ref.~\cite{Shen_MtV})
ranging up to energies being a multitude of the initial temperature. To have an
estimate of the efficiency for transforming the incoming eV laser
photons into hard gamma radiation, we mention that the integrated power
in the $\gamma$ flash is comparable to the total energy of the initial
plasma droplet. Hereby we considered only photons with 
$E_\gamma \ge 1$ MeV. In the considered scenario, the other energy,
after disintegration, resides in collective (radial) flow of electrons and
positrons and soft photons.

\thanks

The authors thank T.E. Cowan and R. Sauerbrey for inspiring debates 
and G. D. Moore for 
supplying the numerical results leading to equation (\ref{cba}).
MGM would like to acknowledge various useful 
discussions with M. H. Thoma and Rajarshi Ray. 
He is also thankful to FZ Dresden-Rossendorf for supporting 
his stay as a FZD-fellow during the course of this work.

\end{document}